\documentclass[12pt,preprint]{JHEP3}
\usepackage{amsmath,amssymb,graphicx,mathrsfs}
\usepackage[latin1]{inputenc}
\usepackage{bm}

\newcommand{\be}{\begin{equation}}
\newcommand{\ee}{\end{equation}}
\newcommand{\bea}{\begin{eqnarray}}
\newcommand{\eea}{\end{eqnarray}}

\title{ Bounds on Black Hole Entropy in Unitary Theories of Gravity }
\author{ Ram Brustein ${}^{(1)}$,  A.J.M. Medved ${}^{(2)}$  \\
(1) Department of Physics, Ben-Gurion University,
    Beer-Sheva 84105, Israel,   \\
(2)    Physics Department,  University of  Seoul,
  Seoul 130-743  Korea  \\
   {\small E-mail: ramyb@bgu.ac.il,\ allan@physics.uos.ac.kr} }
\date{}
\preprint{}
\abstract{
We consider unitary and weakly coupled theories of gravity that extend Einstein gravity and reduce to it asymptotically at large distances. Our discussion is restricted to such theories that, similarly to Einstein gravity, contain black holes as semiclassical states in a range of scales. We show that, at a given scale, the entropy of these black holes  has to be larger than the number of elementary light species in the theory. Our bound follows from the observation that the black hole entropy has to be larger than the product of its mass and horizon radius (in units of Planck's constant divided by the speed of light) and the fact that, for any semiclassical black hole, this product has to be larger than the number of light species. For theories that obey our assumptions, the bound resolves the ``species problem": the tension between the geometric, species-independent nature of black hole entropy and the proportionality of ordinary thermodynamic entropy to the number of species.   We then show that, when black holes in Einstein's theory are compared to those in the  extended theories at a fixed value of mass, the entropy of the Einstein black holes will always be  minimal. Similar considerations are also applied to the entropy density of black branes in anti-de Sitter space.
}
\begin{document}

\section{Introduction}
The entropy of a black hole (BH) depends on a purely geometric property,
its area  in units of Newton's constant \cite{Bek,Haw}. The geometric nature of the BH entropy persists for extensions of Einstein gravity, for which it is given by Wald's  Noether-charge formula \cite{wald1,wald2,wald3}. Then, as shown in  \cite{BGH-0712.3206}, one can always re-express Wald's geometric entropy in the standard area-law form but in  theory-dependent units that are determined by the gravitational coupling. For Einstein's theory, the coupling is simply Newton's constant but, for a general theory, this is not necessarily true. On the other hand, ordinary thermodynamic entropy is proportional to the number of light species $N$.   So, apparently,  if a sufficiently large number of light species existed at a given length scale, then the thermodynamic entropy of the matter that formed a BH larger than this scale would exceed any value of  entropy that does not depend on $N$. Consequently, there appears to be  a severe tension between the purely geometric entropy of  BHs and the second law of thermodynamics. This argument seems to suggest that BHs limit the number of light species and  has been called  ``the species problem" \cite{Jaco}.

In this paper, we consider unitary and weakly coupled theories of gravity that extend Einstein gravity. We further assume that these theories reduce to Einstein gravity asymptotically at large distances. The main differences between Einstein gravity and the extensions under consideration are then expected to occur at small distances.
By unitary, we mean that the theory is free of ghosts below  its ultraviolet (UV) momentum cutoff scale. Since the theory is weakly coupled, we can define the number of light species $N$ more precisely \cite{BDV}. We wish to consider theories that, at an energy scale $\;\Lambda=1/l\;$, have a
finite number $N(l)$ of light species with a mass  $m$ less
than this  scale and with a decay width $\Gamma$ less than their mass; $\;\Gamma<m<\Lambda\;$.  The decay width $\Gamma$ is defined as the inverse of the lifetime of the state, which is the time that it takes for the first transition to a lower energy state via the emission of a light quantum.
The number $N(l)$ includes the graviton and possibly other gravitational degrees of freedom.
Of particular interest is the case that the energy scale $\Lambda$ is at the UV cutoff; $\;\Lambda_{UV}=1/l_{UV}\;$.

We assume that the coupling of all the light species $N(l)$ is such that they can be at thermal equilibrium at (inverse) temperature $\;\beta=l\;$.
We consider only metric theories of gravity
and define the scale $l_{UV}$ for such theories as the scale above which exchanges of metric perturbations in elementary particle processes become strong. Obviously, it follows that, for curvatures less than $1/l_{UV}^2$, gravity is weak and semiclassical. It may well be that Einstein's gravity is modified for scales well above $l_{UV}$; for example,  if large extra dimensions of size $\;R > l_{UV}\;$ exist.

Let us recall the properties of any semiclassical BH, as described in \cite{BDV}, since these are central to the following discussion. We consider, as in \cite{BDV}, neutral, static and non-rotating BHs  and recall that these can be described in terms of three parameters: the mass $M$, the Schwarzschild radius $R_S$ and the inverse temperature $\beta=1/T$. By the no-hair theorem \cite{nohair},  these parameters must be related in a simple way for any theory of gravity; however, the precise relation is theory dependent. The following conventions are used throughout:
$\hbar$, $k_B$, $c = 1$, the number of dimensions of the spacetime is $\;D=d+1=n+2\geq 4\;$ but we  mostly limit ourselves to $D=4$.  As our emphasis is on the parametrical behavior of quantities,
numerical factors of order unity are consistently neglected.
The subscripts $E$ and $X$ are always used to denote
quantities in Einstein's theory and a generic extension, respectively. If no subscript appears on a certain quantity then it is relevant to all theories; for example, $R_{S;E}$ is the Schwarzschild radius of a BH in Einstein gravity,  $R_{S;X}$ is the Schwarzschild radius of a BH in an extended gravity, and $R_{S}$ denotes a Schwarzschild radius of a BH in either. For simplicity, we mainly consider static ``Schwarzschild-like'' BHs in an asymptotically flat spacetime. Analogous bounds can typically be formulated for more complicated scenarios.

As in \cite{BDV}, we assume that semiclassical BHs are
\begin{list}{$i)$}
\item Black bodies,
\end{list}
\begin{list}{$ii)$}
\item
Classically black.
\end{list}
\ \\ Assumption $i)$ can be stated as
$$
-\frac{dM}{dt}\;=\; N(\beta) \beta^{-4} R_S^2 \; .
$$
Here, $N(\beta)$ is the number of light species into which the BH can decay.
Assumption $ii)$ implies that the quantum wavelength of particles emitted by the BH is at least as large as  $R_S$. Since, according to assumption $i)$, the BHs are  black bodies, this implies that the energy of emitted quanta $R_S^{-1}$ also  bounds the BH temperature, so
\begin{equation}
\label{rsbeta}
R_S/\beta \;\le\; 1\;.
\end{equation}

Following \cite{Dv1}-\cite{Dv5} and \cite{BDV}, we further assume that, for semiclassical BHs, the size and inverse temperature decrease slower than the speed of light:
$$(a)\ - \frac{dR_{S}}{dt}\; < \; 1\;, $$
$$(b)\ \ \ - \frac{d\beta}{dt}\; <\; 1\; .$$ Additionally, the fraction of the mass loss of the BH has to be small during both the thermal and the light crossing time scales:
$$(c)\
-\frac{R_S}{M} \frac{dM}{dt}\;<\; 1\; ,$$
$$\ (d)\
-\frac{\beta}{M} \frac{dM}{dt}\;<\; 1\; .$$
Finally, we require that the BH be metastable or
$$(e)\hspace{.5in}
\frac{\Gamma}{M}\; < \; 1\; .
$$
The width $\Gamma$ is defined the same as for the elementary species: the inverse of the time it takes for the first transition from the BH to a lower energy state by the emission of a light quantum.
In the preceding discussion, we have ignored  grey factors and  numerical factors related to the statistics of the species. None of the decay channels of the BH are expected to be parametrically suppressed at energies $\sim 1/\beta$, which is the main energy range of the BH emission.   Since the BHs are assumed to be black bodies, it follows that  $N(\beta)$ is equal to the number of species that can be in thermal equilibrium at (inverse) temperature $\beta$.
In \cite{BDV}, it was shown that  BHs which satisfy all the above assumptions obey the important inequality
\begin{eqnarray}
\label{bounds1b}
R_S M \;>\; N(\beta)\;.
\end{eqnarray}

As argued  by Dvali and others
\cite{Dv1}-\cite{Dv5}, through the implementation
of non-perturbative consistency arguments,
the radial size of a black hole for Einstein's theory is limited by
the minimal scale $l_{UV;E}$ such that
$\;l^2_{UV;E}= G_E N_E(l_{UV;E})\;$. Physically, $l_{UV;E}$ is the effective UV
cutoff for the theory, below
which semiclassical gravity can no longer be trusted.
With this scale as a lower bound on $R_E$,
along with  the standard area--entropy law $\;S= R_{S;E}^2/G_E\;$,  the inequality
\be
S_{E} \;>\; N_E(l_{UV;E})\;
\label{nbound1}
\ee
then follows.

\section{The number of species bounds black hole entropy}

We wish to generalize bound~(\ref{nbound1}) and find out whether it is also valid for extensions of Einstein gravity.

Let us relate the BH entropy $S_X$  to its mass $M_X$. Using our assumption that semiclassical BH's exist for a range of masses, we can then consider the situation of a BH changing its mass, size and temperature such that  the resulting state is another semiclassical BH. This allows us to invoke the first law of BH mechanics in its integral form.

Recalling the  differential  form  of the first law
of BH mechanics, we have
\begin{equation}
\;dS_X \;=\; \beta_X(M_X)dM_X\;;
\label{frstlaw}
\end{equation}
from which  the BH entropy is given by
\begin{equation}
S_X \; = \;\int \beta_X(M_X)dM_X\;.
\label{int-ent}
\end{equation}
Since bound~(\ref{rsbeta}) is valid for all values of the mass,  $\;\beta(M_X)\geq R_{S;X}(M_X)\;$ and thus
\begin{equation}
\int \beta_X(M_X)dM_X \; \geq \;\int R_X(M_X)dM_X \;.
\label{int-rm}
\end{equation}

For a Schwarzschild BH in  Einstein's theory, bound~(\ref{int-rm})  is saturated, $\;\beta_E(M_E)= R_{S;E}(M_E)\;$, leading to the integrated first law $\;S_E=R_{S;E} M_E\;$.
Meanwhile, for any weakly coupled extension, $\;R_{S;X}(M_X)=M_X[1 +I_X(M_X)]\;$
in Planck units, where the ``correction'' $M_XI_X$ is known to be
positive definite and of order unity (see Eq.~(13) of \cite{BDV} and
the surrounding discussion). It  follows that
\begin{equation}
\;\int R_{S;X}(M_X)dM_X \;=\; R_{S;X} M_X\;,
\label{inrxmx}
\end{equation}
up to a parametrically small correction (the difference between
$\int I_X M_X dM_X$
and $I_XM_X^2$, which is generically of order $1$)
 and a constant of integration. We
deal with the latter by extrapolating the semiclassical regime down to
$M_X=0$ and then
 making the natural assumption that, just like for Einstein gravity, the
integration constant is zero. That is, a vanishing mass means  a  non-existent BH horizon with the  corresponding limit $\;R_{S;X}(M_X\to 0)\to 0\;$
being well defined.

Combining Eqs.~(\ref{int-ent}), (\ref{inrxmx}) and bound~(\ref{int-rm}), we  obtain the entropy bound
\be
S_X\; \geq \;  R_{S;X} M_X \; .
\label{massbound}
\ee

Then, combining inequality~(\ref{bounds1b}) and bound~(\ref{massbound}), we arrive at the desired inequality
for any semi-classically allowed value of $\beta_X$:
\be
S_X  \; > \; N_X(\beta_X) \;.
\label{esen}
\ee
Since inequality~(\ref{esen}) is valid for all scales, it is also valid for $\;\beta_X\sim l_{UV;X}\;$, and so
\be
S_X \; >\; N_X(l_{UV;X})\;,
\label{nbound2}
\ee
which is the obvious analogue of Eq.~(\ref{nbound1}).

We expect that $N_X(l)$ will monotonically increase with increasing energy scale
or  decreasing $l$. Hence, it is expected that the strongest possible  bound on the entropy is
when $\beta_X$ reaches the UV cutoff $l_{UV;X}$ for the generic theory;
that is, the inequality~(\ref{nbound2}) is the  most restrictive one.

\section{The entropy of black holes in Einstein gravity is minimal}

We can make further use of bound~(\ref{massbound})
to show that $\;S_X(M)\geq S_E(M)\;$ for a fixed mass $M$.

Our other key input is that, for a unitary extension of Einstein gravity,
any gravitational coupling {\em in the vicinity of the horizon}
can only increase from
its Einstein value. To understand this point, let us consider
small perturbations  of the background (Schwarzschild-like)
metric near the horizon:
$\;g_{\mu\nu}\rightarrow g_{\mu\nu} + h_{\mu\nu}\;$ with $\;h_{\mu\nu}\ll 1\;$.

Since the perturbations are small,
the one-particle exchange approximation can be invoked to
determine the coupling of the gravitons. For those near
the BH horizon, it is
appropriate to use a flat-space form of the 1PI propagator.
Indeed,
in spite
of the curvature of the  BH background, a particle near the horizon will
 ``observe'' an effective geometry
that is two dimensional and conformally flat.
This becomes evident, as we next discuss,
from the near-horizon form of the Klein--Gordon equation.

For a Schwarzschild-like BH solution  the metric can  be expressed
in the form
$\;ds^2=-f(r)dt^2+\frac{1}{g(r)}dr^2+r^2d\Omega_n^2\;,$
where $f(r)$ and $g(r)$ have simple zeroes at the
horizon $\;r=R_S\;$, while $\;\partial_rf(r)|_{r=R_S}\;$ and $\;\partial_rg(r)|_{r=R_S}\;$
are both finite and non-vanishing.
It is convenient to introduce the usual radial ``tortoise''
coordinate $r_{*}$ such that $\;dr_{*}\equiv dr/\sqrt{fg}\;$; then
$\;ds^2=f(r_{*})\left[-dt^2+dr_{*}^2\right]+r^2d\Omega_n^2\;$.
Let us, for simplicity, consider the Klein--Gordon equation for
a scalar (the results also apply  to gravitons):
$\;(\Box \phi - m^2)\phi =0\;$ or
\be
\left[\;-\frac{1}{f}\partial_t^2\;+\;\frac{1}{r^2 f}
\partial_{r_{*}}\left(r^2\partial_{r_{*}}\right)\;
+\; \frac{1}{r^2}\partial^2_{\Omega}\;-\;m^2\;\right]\phi=0\;,
\ee
where $\partial^2_{\Omega}$  represents the
sum over angular derivatives.
Multiplying through by $f$ and using the defining
relation for $r_{*}$,  we  obtain
\be
\left[\;-\partial_t^2\;+\;\frac{\sqrt{fg}}{r}
\partial_{r_{*}}\;+\;\partial^2_{r_{*}}\;
+\; \frac{1}{r^2}f\partial_{\Omega}^2\;-\;f m^2 \;\right]\phi=0\;.
\label{stuff}
\ee
Taking the near-horizon limit, we find that
Eq.~(\ref{stuff}) reduces to
\be
\left[\;-\partial_t^2\;+\;\partial^2_{r_{*}}\;
\right]\phi=0\;.
\ee
That is,  given a scattering process in the vicinity of the horizon,
any of the involved particles are essentially massless and effectively
``perceive'' a two-dimensional, flat spacetime.

Let us now discuss a two-graviton scattering process in a flat spacetime.
For Einstein gravity, only massless
spin-$2$ gravitons are exchanged, but gravitons can,
for a general theory, be  either   massless or
massive and either spin-$0$ or spin-$2$. Particles of
any other spin, in particular vectors, can not
couple linearly  to a  conserved source and
may therefore  be ignored.
Consequently, the 1PI graviton  propagator
$\;[{\cal D}(q^2)]^{\ \nu \; \beta}_{\mu \; \alpha}\equiv \langle h_{\mu}^{\ \nu}(q)
h_{\alpha}^{\ \beta}(-q)\rangle\;$
has the following irreducibly decomposed form
\cite{Zakharov}-\cite{Dvali2}:
\bea
[{\cal D}(q^2)]^{\ \nu \; \beta}_{\mu \; \alpha} \;&=&
\; \left(\rho_E(q^2) + \rho_{NE}(q^2)\right)
\left[\delta^{\ \beta}_{\mu}\delta^{\ \nu}_{\alpha}-
\frac{1}{2}\delta^{\ \nu}_{\mu}\delta^{\ \beta}_{\alpha}\right] \frac{G_E}{q^2}
\nonumber \\
&+&\!\!\sum_i\rho^i_{NE}(q^2)\left(\delta^{\ \beta}_{\mu}\delta^{\ \nu}_{\alpha}-
\frac{1}{3}\delta^{\ \nu}_{\mu}\delta^{\ \beta}_{\alpha}\right)\frac{G_E}{q^2+m_i^2}
\nonumber \\
&+&\!\!\sum_j {\widetilde\rho}^{\; j}_{NE}(q^2)\;
\delta^{\ \nu}_{\mu}\delta^{\ \beta}_{\alpha}\;
\frac{G_E}{q^2 +{\widetilde m}_j^2}\;.
\label{decomp}
\eea
Here, the positive number  $\;q^2=-q^{\mu}q_{\mu}\;$ is the flat-spacetime (spacelike) momentum,
the propagator is evaluated in the vacuum state and
$G_E$ is the $D$-dimensional Newton's constant. Also, the
Einstein  and ``Non-Einstein'' parts of the gravitational couplings $\rho$
are indicated by the subscripts $E$ and $NE$.
We have distinguished between the
massless spin-$2$ particles,
massive spin-$2$ particles with mass $m_i$ and
scalar particles with mass ${\widetilde m}_j$.
The $\rho$'s  are dimensionless couplings that generally
depend  on the energy scale $q$; in particular,  $\;\rho_E(0)=1\;$ and $\;\rho_{NE}(0)=0\;$.

Although valid for a flat spacetime, Eq.~(\ref{decomp}) is  also
applicable to a scattering near a BH horizon where the spacetime
is, as shown above, effectively two-dimensional and flat. The internal states are limited to massless
particles propagating strictly in the radial direction, so that
all the $m$'s  vanish and $\;q^2\to q^2_{2D}=q^2_t-q^2_{r_{*}}\;$.
Hence,
\bea
[{\cal D}_{r=R}(q^2=q^2_{2D})]^{\ \nu \; \beta}_{\mu \; \alpha} \;&=&
\; \left(\rho_E(q^2_{2D}) + \rho_{NE}(q^2_{2D})\right)
\left[\delta^{\ \beta}_{\mu}\delta^{\ \nu}_{\alpha}-
\frac{1}{2}\delta^{\ \nu}_{\mu}\delta^{\ \beta}_{\alpha}\right] \frac{G_E}{q^2_{2D}}
\nonumber \\
&+&\!\!\sum_i\rho^i_{NE}(q^2_{2D})\left(\delta^{\ \beta}_{\mu}\delta^{\ \nu}_{\alpha}-
\frac{1}{3}\delta^{\ \nu}_{\mu}\delta^{\ \beta}_{\alpha}\right)\frac{G_E}{q_{2D}^2}
\nonumber \\
&+&\!\!\sum_j {\widetilde\rho}^{\; j}_{NE}(q^2_{2D})\;
\delta^{\ \nu}_{\mu}\delta^{\ \beta}_{\alpha}\;
\frac{G_E}{q^2_{2D}}\;,
\label{decomp2}
\eea
whereas $\;[{\cal D}_{r=R}(q^2\neq q^2_{2D})]^{\ \nu \; \beta}_{\mu \; \alpha} \;=\;0\;$.

Eq.~(\ref{decomp2}) describes the annihilation
of a graviton at one point in the near-horizon region followed by its creation at another point
in the same region, with virtual processes taking place in between.    It is important to emphasize that the
$D$-dimensional origin of the propagator is still encoded in Eq.~(\ref{decomp2}). The external indices $\mu$,$\nu$,$\alpha$,$\beta$ can take on any of the original  $D$-dimensional values. Additionally, each $D$-dimensional field leads to many two-dimensional fields having the same two-dimensional momentum but with a slightly different angular profile. So that, even though the internally scattered particles are
massless and restricted to radial motion, the $\rho$'s still  keep track of the original $D$-dimensional fields.

For a unitary theory, all of the couplings
must remain positive at all energy scales \cite{Dvali1};
meaning that {\em the propagator can  only increase}
relative to its Einstein value. Let us recall the Kallen--Lehmann representation
of the 1PI propagator (see, {\it e.g.}, Ch.~12 of \cite{srednicki}).
We observe that the $\rho$'s are spectral densities
that can be microscopically evaluated with the insertion of a complete
set of single- and multi-particle states; schematically,
$\;\rho=\sum_n<0|h|n><n|h|0>\;$. If all such inserted states have a
positive norm, there can only be positive contributions
to a given $\rho$. By the same token, a generalized theory that
extends Einstein's can only make a positive contribution
to any  Einstein spectral density or gravitational coupling.
To rephrase, any additional degree of freedom will open up  additional
intermediating attractive channels.
The extra channels can only act to increase  the
overlap between two given graviton states
at two different points, thus  leading to an increase in the associated gravitational coupling.

Let us recall that, for  a generalized  gravity theory, the BH entropy is given by Wald's Noether-charge formula \cite{wald1,wald2,wald3}, which can then be re-expressed \cite{BGH-0712.3206} as an area law
with a theory-dependent gravitational coupling $G_X$.
For Einstein's theory, $\;G_X= G_E\;$ is simply Newton's constant. More
generally, $G_X$  determines the strength of the $r$,$t$-polarized gravitons in the vicinity
of the horizon and,
thus,  can also be  extracted from  the near-horizon propagator
for the $h_{rt}$ gravitons \cite{BGH-0712.3206,BM-0808.3498}.
Then, from Eq.~(\ref{decomp2}), it follows that
$\;G_X\geq G_E\;$ must hold true at any given energy scale.
We cannot, however,  directly apply
Eq.~(\ref{decomp2}) to determine how the entropy $S_X$ compares
to the  Einstein value (as was done previously for the shear viscosity
of a brane theory via the $h_{xy}$ gravitons \cite{BMlatest}).
This is because, for a BH (and unlike for
a black brane),  the  horizon radius also depends on
the strength of the coupling.

We would like to compare the generalized entropy $S_X$ to its
Einstein counterpart $S_E$ at a fixed value of the BH mass, as
a microcanonical calculation is the most appropriate
one for a static BH. However, the mass is defined at infinity, whereas we require a quantity that is determined at the horizon.
For $D=4$, a natural choice
is  to use  $\;{\widetilde M}_X\equiv R_X/G_X$\;. For Einstein gravity, $\;{\widetilde M}_E=M_E\;$.

The Wald prescription tells us that
the corresponding BH entropy is
$\;S_X= R_{S;X} {\widetilde M}_X =R_{S;X}^2/G_X\;$ or
$\;S_X=  G_X {\widetilde M}_X^2\;$.
Let us first choose  $\;{\widetilde M}_X={\widetilde M}_E\;$
and take the appropriate ratio to obtain
\be
\frac{S_X}{S_E}_{|\;{\widetilde M}_X = {\widetilde M}_E}\;=\; \frac{G_X}{G_E}\;\geq \; 1\;.
\label{ratio1}
\ee

Additionally, using  bound~(\ref{massbound}) and the fact that
$\;S_X=R_{S;X}{\widetilde M}_X \;$,  we obtain $\;{\widetilde M}_X \geq M_X\;$.
Then, recalling that  $\;{\widetilde M}_E=M_E\;$, we have
\be
\frac{S_X}{S_E}\;=\;\frac{G_X {\widetilde M}^2_X}{G_E {\widetilde M}^2_E}
\;=\; \frac{{\widetilde M}_X^2}{M_X^2}\frac{G_X M^2_X}{G_E M^2_E} \;.
\ee
Finally, fixing the mass $\;M_X=M_E\;$, we arrive at
\be
\frac{S_X}{S_E}_{|\;{M}_X = { M}_E}\;=\; \frac{{\widetilde M}_X^2}{M_X^2}\frac{G_X }{G_E}
\;\geq \; \frac{G_X}{G_E}\;\geq\; 1\;.
\label{ent-ratio}
\ee

Previously we have found that $\;S_X > N_X\;$ and
$\;S_E > N_E\;$. Can we now conclude that
$\;
N_X(l_{UV;X})\geq N_E(l_{UV;E})\;
$?
It is already known that $\;l_{UV;X}\geq l_{UV;E}\;$ \cite{BDV}.
Moreover,  at any given  scale, a  weakly coupled unitary extension of Einstein's theory has at least as
many  degrees of freedom as that of Einstein gravity.
That is, $\;N_X(l)\geq N_E(l)\;$ can be expected to
hold for any $l$ within the semi-classical regime
of both theories.
It thus follows that $\;N_X(l_{UV;X})\geq N_E(l_{UV;X})\;$ is in fact valid, where
we  have used the UV scale $l_{UV;X}$
in the arguments for both $N_X$ and $N_E$.

\section{The entropy density of black branes in AdS}

Some of  the preceding analysis can be extended  to the case of an asymptotically anti-de Sitter (AdS) brane
theory, where the quantity of interest is the entropy density
of the black brane. The near-horizon form of an AdS brane metric is
essentially  the
same as that of a BH in an asymptotically flat spacetime. Hence,
we can apply the same reasoning
as before and again  use the flat-space form of the 1PI propagator
or Eq.~(\ref{decomp2}). In this context,  the stable tachyons of
AdS space do not pose a special problem. Since masses come attached with a
factor of $f$ ({\it cf}, Eq.~(\ref{stuff})), any finite-mass  tachyon
will also behave as a massless species.
As it is well known, the condition of stability  restricts the magnitude of any tachyon mass  to
be finite and small, $\;|m^2|\lesssim d/L^2\;$ \cite{maldacena2},
where $L$ is the AdS radius of curvature.

We wish to compare the entropy density $s_X$
for a generalized theory to the entropy density $s_E$ of
Einstein's theory, and do so
at a common value for the Hawking temperature, $\;T_X=T_E\equiv T\;$, with all other charges fixed. This canonical
framework is the obvious analogue of comparing BHs at fixed mass,
given that the energy of a brane is infinite. Then,
since $\;s_E\sim T^n/G_E\;$ and $\;s_X\sim T^n/G_X\;$,  any difference in the densities must be due solely to the  gravitational couplings. Using our previous arguments,
we expect that $\;G_X\geq G_E\;$, and so an ``inverted bound'' should follow:
\be
s_X \;\leq\; s_E\;.
\label{densbound}
\ee
However, this is not the story in its entirety. A precursor is the fact that
bound~(\ref{densbound}) is  not invariant under field redefinitions.

To compute $G_X$, we can again make use of the Wald formalism
\cite{wald1,wald2,wald3}
as prescribed
in \cite{BGH-0712.3206}.
For this coupling to deviate physically from $G_E$,  it is necessary
that the Lagrangian depends explicitly  and non-linearly on the Riemann tensor in a polarization-dependent form. Any deviation in $G_X$ from $G_E$  that is independent of the graviton polarization is not  physically meaningful, as  $G_X$ could always be recalibrated to be equal to $G_E$, with all other couplings then being rescaled by the same amount.

However, theories that depend on higher powers of the Riemann tensor lead to, generically, equations of motion with more than two time derivatives and, therefore,  to propagating ghosts. Well-known non-generic exceptions are the Kaluza--Klein  scenarios, whereby $G_X$  may change from $G_E$ but in  such a way that
$s_X=s_E$ is always  maintained \cite{BMlatest}. In the Kaluza--Klein scenarios one needs to keep an infinite number of higher derivative terms.

The only  consistent unitary extensions of Einstein gravity with a finite number of additional higher-curvature terms are those of the Lovelock class \cite{LL,Zwi,myers}.
But, for any Lovelock brane theory, one finds that $\;G_X=G_E\;$ must be true!
To demonstrate this last point, let us first express the brane metric
in a generic form, assuming only translational invariance
on the brane and a static spacetime:
\be
ds^2= -a(r)f(r)dt^2+\frac{1}{b(r)}\frac{1}{f(r)}dr^2+\sum_{i=1}^n c(r)dx_i^2\;.
\label{brane}
\ee
Here, $f(r)$ has a simple zero at the horizon $R_S$; otherwise, the  radial functions $a(r)$, $b(r)$, $c(r)$ and their derivatives (including
that of $f(r)$) are positive and regular in
the exterior region of the spacetime.

For a Lovelock theory, the non-Einstein contributions
to the Wald entropy are strictly constructed out of contractions
of $4$-index Riemann tensorial components  that do {\em not} carry any
$r$ and/or $t$ indices \cite{Jac,Vis}. That is, these contributions will be
made up only of contractions of ${\cal R}_{x_{i}x_{j}x_{i}x_{j}}$
(with $\{i\neq j\}=\{ 1,2,\ldots,n\}$) and permutations thereof.
Now consider that
$\;{\cal R}^{x_{i}}_{\;\;x_{j}x_{i}x_{j}}=\Gamma^{x_i}_{x_i r}\Gamma^{r}_{x_jx_j}
\sim f b\left[\partial_r c \right]^2/c\;$. At the horizon, $f$ is vanishing,
while $b$, $c$ and $\partial_r c$ are all regular --- meaning that
any  of these tensorial components and, hence, the non-Einstein contributions
to the Wald entropy must be vanishing for this Lovelock brane theory.

As explained above, $\;s_X/s_E=G_E/G_X\;$ when evaluated at
the same value of temperature. For Lovelock theories,
the implication is clear:
\be
\frac{s_X}{s_E}= 1\;.
\label{ratio2}
\ee
Note that this equality of entropy densities is applicable to both
the gravity theories and their
field-theory duals \cite{KSS-hep-th/0309213,SS-0704.0240}.

Equation~(\ref{ratio2}) can be used to directly verify the  so-called KSS bound
\cite{KSS-hep-th/0405231}. This bound asserts
that $\;\eta/s\geq1/4\pi\;$ --- where $\eta$ is the shear viscosity
of the brane and its field-theory dual alike    ---
and can be elegantly restated as
$\;\eta_X/s_X\geq \eta_E/s_E\;$.
The key point here is that the factors of temperature now cancel
out of this ratio. And so  this is really a direct comparison
of gravitational couplings, which are generally different
for the different polarizations involved
(with $s$ and $\eta$ implicating the $r$,$t$- and $x_{i}$,$x_{j}$-polarized gravitons
respectively \cite{BM-0808.3498}).
In a previous work \cite{BMlatest}, we have used
Eqs.~(\ref{decomp}~,~\ref{decomp2}) to argue that $\;\eta_X\geq \eta_E$\;.
(Note that the horizon is the relevant surface for calculations of $\eta$,
even for the dual theory \cite{Cai}.)
Taken together with either  Eq.~(\ref{densbound}) or (\ref{ratio2}), the KSS bound immediately
follows.

\section{Summary and Conclusion}

To summarize, we have demonstrated two novel bounds for the
entropy of a semiclassical BH in a unitary and weakly coupled
extension of Einstein gravity.  We have established that, at any applicable energy scale up to the UV momentum cutoff scale, this entropy must be larger than the number of elementary light species in the theory at the given scale.
Then, we have verified that, for different theories  compared at a fixed value of mass, the entropy will always be minimal for Einstein gravity. Some of  the same basic principles were also applied to the entropy density of a generalized AdS brane theory.

The bound  $S > N$  provides a  resolution to the so-called species problem. It places a fundamental and universal limit on the entropy of a semiclassical BH in terms of the number of light particle species. The bound implies that the BH entropy apparently does ``know'' about the various particle species but, at the same time, is able to maintain its status as a purely geometric quantity. Our considerations suggest the likely route for resolving the tension between the two different facets of the BH entropy: In a theory with an excessively large number of elementary fields, BHs that have less than the required entropy are not stable enough to be a part of the semiclassical spectrum. They either decay too quickly to some other states, either elementary particles or BHs after they form, or do not have enough time to form at all. It would be interesting to understand in detail the dynamical physical mechanisms that act to enforce the bound $\;S>N\;$  when it is near saturation.

The fact that the entropy of BHs in Einstein gravity is minimal compared to any BH of the same mass in an extended theory  implies that, when one considers bounds on the entropy of BHs and the significance of such bounds, then it is enough to establish them for Einstein's theory. Since Einstein gravity is much simpler to analyze than most of its extensions, this bound is technically very useful. It would be interesting to determine explicitly whether this bound applies to string theoretical BHs. Our results suggest that it does. In \cite{gomezdvali}  the entropy bound is in fact used to constrain the effective number of species in string theory.

\section{Acknowledgments}

The research of RB was supported by The
Israel Science Foundation grant no 470/06.
The research of AJMM is supported by the University of Seoul.
AJMM thanks  Ben-Gurion University for their hospitality during his visit.

\end{document}